\documentclass[aps,pra,twocolumn,superscriptaddress,showpacs]{revtex4}

\usepackage{amsmath,amsfonts,amssymb,bbm,citesort,graphicx}

\newcommand{\eqn}[1]{\begin{equation} #1 \end{equation}} 
\newcommand{\aln}[1]{\begin{align} #1 \end{align}}       
\newcommand{\mul}[1]{\begin{multline} #1 \end{multline}} 
\newcommand{\mc}{\mathcal}                               
\newcommand{\mbf}{\mathbf}                               
\newcommand{\bs}{\boldsymbol}                
\newcommand{\eq}[1]{(\ref{#1})}              
\newcommand{\pd}{\partial}                   
\newcommand{\wtilde}{\widetilde}             
\newcommand{\wbar}{\overline}                
\renewcommand{\l}{\left}                     
\renewcommand{\r}{\right}                    

\begin{document}


\title{Diagrammatic semiclassical laser theory}


\author{Oleg Zaitsev}
\email[E-mail: ]{oleg.zaitsev@uni-bonn.de}
\affiliation{Physikalisches Institut der Universit\"at Bonn, Nu{\ss}allee 12,
             53115 Bonn, Germany}

\affiliation{Fachbereich Physik, Universit\"at Duisburg-Essen,
             Lotharstra{\ss}e~1, 47048 Duisburg, Germany}

\author{Lev Deych}
\email[E-mail: ]{lev.deych@qc.cuny.edu}
\affiliation{Physics Department, Queens College of City University of
             New York, Flushing, NY 11367, U.S.A.}


\begin{abstract}

We derive semiclassical laser equations valid in all orders of nonlinearity.
With the help of a diagrammatic representation, the perturbation series in
powers of electric field can be resummed in terms of a certain class of
diagrams. The resummation makes it possible to take into account a weak effect
of population pulsations in a controlled way, while treating the nonlinearity
exactly. The proposed laser theory reproduces the all-order nonlinear equations
in the approximation of constant population inversion and the third-order
equations with population-pulsation terms, as special cases. The theory can be
applied to arbitrarily open and irregular lasers, such as random lasers.

\end{abstract}


\maketitle


\section{Introduction}

Interest in random lasers with coherent feedback~\cite{frol99,cao99} and lasers
based on chaotic microresonators without mirrors~\cite{fang07} revealed a
number of shortcomings of conventional laser theory~\cite{sarg74,hake84} that
complicated its application to such systems. Among the properties that
characterize random and chaotic lasers are strong openness, irregular spatial
dependence of the refractive index, and, possibly, nontrivial shape of the
resonator. Lasing in these systems can be accompanied by strong coupling
between the modes, which requires a more careful treatment of nonlinear effects
than is necessary for regular lasers.

An essential part of a laser description is the choice of an appropriate basis
of normal modes in which electromagnetic field and other system functions can
be expanded. In an open system it is not possible to define a Hermitian
eigenvalue problem whose eigenfunctions would form an orthogonal basis.
Instead, one has to introduce a biorthogonal system of so called quasimodes as
left and right eigenfunctions of a non-Hermitian operator. A number of methods
to construct quasimodes were discussed in the literature. Among the earliest
are the Fox-Li modes~\cite{fox68,sieg89a,sieg89b} which are useful in
resonators with a preferred propagation direction and clearly defined
transverse plane. In a more general setting, there were attempts to use
solutions of the wave equation satisfying outgoing boundary conditions at
infinity (Siegert-Gamow boundary conditions), with complex eigenfrequencies
corresponding to scattering resonances~\cite{mois98,chin98,dutr00}. However,
these modes diverge at infinity, which makes it problematic to use them as a
basis~\cite{mois98}, while some ways around this problem have been discussed in
Refs.~\cite{chin98,dutr00}. Another possibility is to use the so called
system-and-bath type approaches~\cite{vivi03,vivi04}, where cavities are
described by an orthogonal system of wavefunctions of a Hermitian problem with
an independent sets of modes introduced for outside of the resonator. The
openness of the cavity in this approach is reproduced by introducing coupling
between discrete set of inside modes and the continuous spectrum of outside
modes. This way, the modal expansion of the field becomes possible both inside
and outside of the cavity.

In the present work we use a different approach based on keeping the spectral
parameter of the outside outgoing field real, while making inside field to
satisfy continuity conditions at the boundary of the cavity. This approach also
results in the discrete spectrum of the cavity field with complex-valued
frequencies, but, since the outside field is forced to depend on real spectral
parameter, it does not diverge and is characterized by a constant flux. This
approach was used in Ref.~\cite{mors53} for special kind of vibration problems
and was adapted for lasers in Ref.~\cite{ture06}, where respective modes were
dubbed the ``constant-flux'' (CF) modes. One can introduce two adjoint
biorthogonal systems of CF modes, which can be used to represent field inside the
cavity.

In the standard semiclassical laser theory~\cite{sarg74,hake84} lasing modes are
usually taken to coincide with quasimodes of the respective cavities, while
their amplitudes and frequencies are found from equations based on perturbation
expansions containing terms linear and cubic in the field.  In random lasers
this picture needs to be revised. First, it was shown~\cite{deyc05a,deyc05b}
that normal modes in the presence of gain differ from the passive modes even in
the linear approximation if the refractive index and/or unsaturated population
inversion are nonuniform. Second, it was realized~\cite{misi98,hack05,zait09}
that self- and cross-saturation coefficients before the cubic terms can have
different statistical properties in different systems, leading to different mode
statistics. Finally, it was pointed out~\cite{ture06,ture08} that nonlinear
effects can significantly contribute to modification of the lasing modes
compared to those of the empty cavity. A theory suggested in
Refs.~\cite{ture06,ture08}  allowed for self-consistent calculations of not only
lasing frequencies but also of the spatial distributions of the respective
modes.  Neglecting pulsations of the population inversion the authors of
Refs.~\cite{ture06,ture08} were able to derive equations for field amplitudes
and frequencies beyond the usual third-order approximation.

In the present work we also generalize the conventional laser theory, but,
unlike the approach of Refs.~\cite{ture06,ture08}, we do not begin by
introducing a special approximation for population inversion. Instead, we carry
out the perturbation expansion up to the infinite order in the field keeping
all the terms which do not have fast temporal oscillations. This also includes
a part of population-pulsation contributions which is consistent with the
slowly-varying-envelope approximation. The classification of the resulting
terms becomes possible due to a special diagram technique developed to
represent the terms of the expansion. This technique, however, differs
significantly from usual Feynman diagrams widely used in solid state and
high-energy physics, because we have to deal with terms of ever increasing
degree of nonlinearity. Thus, diagrams in our technique are not used to
literally represent each term of the expansion, but play a more limited role,
as a tool assisting in the classification of the terms. Nevertheless, this
technique allows for standard division of diagrams into connected and
disconnected, with disconnected diagrams submitting to easy resummation in
terms of only connected ones. The connected diagrams can be classified
according to the order of magnitude of the population pulsations. The resulting
laser equations generalize the nonlinear equations of
Refs.~\cite{ture06,ture08} in two respects. First, equations derived in this
paper are dynamic, allowing for studying time dependence of the amplitudes,
while the equations of Refs.~\cite{ture06,ture08} can only describe stationary
lasing output. Second, our equations incorporate terms responsible for the
population-pulsation contribution in all orders of the perturbation theory; the
equations of Refs.~\cite{ture06,ture08} are reproduced in our approach if only
the lowest-order diagram is taken into account.

The structure of our paper is as follows. In Sec.~\ref{sec:qm} we recall the
definition and properties of the constant-flux quasimodes of open system.
Standard semiclassical laser equations are written in Sec.~\ref{sec:sle} in
frequency representation for later convenience. Coupled equations for electric
field, polarization, and population inversion are reduced to equations for the
field alone in Sec.~\ref{sec:all} using infinite-order perturbation theory. In
Sec.~\ref{sec:dt} we formulate the diagrammatic technique and resum the
perturbation expansion in terms of connected diagrams. In Sec.~\ref{sec:lim} we
reproduce the results of linear theory, third-order theory with
population-pulsation terms, and all-order nonlinear theory in the
constant-inversion approximation. Finally, we write corrections to the
all-order theory that are of the first order in the population pulsations.

\section{Constant-flux quasimodes
\label{sec:qm}}

We consider an open system defined by real dielectric constant~$\epsilon (\mbf
r)$, with $\epsilon = 1$ outside of the system's boundary. In the Coulomb gauge
$\nabla \cdot \l[\epsilon (\mbf r)\,  \mbf E (\mbf r, t) \r] = 0$, an electric
field $\mbf E (\mbf r, t)$ is governed by the wave equation
\eqn{
  \epsilon (\mbf r)\, \frac {\pd^2} {\pd t^2} \mbf E + \nabla \times
  \l(\nabla \times \mbf E \r) = 0,
\label{weqpas}
}
where Gaussian units with the velocity of light in vacuum $c = 1$ are used.

In the absence of gain and absorption the field will decay in time due to the
openness. It is convenient to represent the decaying field as a superposition
of certain normal modes, the quasimodes, that have only
outgoing components outside the system. These modes can be constructed as
families of constant-flux (CF) states~\cite{ture06} $\bs \psi_k (\mbf r,
\omega)$ depending on a real continuous parameter~$\omega$. Explicitly, the CF
modes satisfy the differential equation
\eqn{
  \nabla \times \l[ \nabla \times \bs \psi_k (\omega) \r] = \omega^2\, \bs
  \psi_k (\omega)
\label{cfout}
}
in the exterior of the cavity with the outgoing-wave boundary conditions at
infinity. Inside the system, the same state satisfies a different equation:
\eqn{
  \frac 1 {\sqrt{\epsilon (\mbf r)}} \nabla \times \l[ \nabla \times \frac
  {\bs \psi_k (\omega)} {\sqrt{\epsilon (\mbf r)}} \r] =
  \Omega_k^2 (\omega)\, \bs \psi_k (\omega).
}
For each $\omega$, the complex eigenfrequency $\Omega_k (\omega)$ is quantized,
as the eigenfunctions are required to match smoothly at the interface.

Conjugate wavefunctions $\bs \phi_k (\mbf r, \omega)$ obey Eq.~\eq{cfout}
outside with the incoming-wave boundary conditions and the equation
\eqn{
  \frac 1 {\sqrt{\epsilon (\mbf r)}} \nabla \times \l[ \nabla \times \frac
  {\bs \phi_k (\omega)} {\sqrt{\epsilon (\mbf r)}} \r] =
  [\Omega_k^2 (\omega)]^*\, \bs \phi_k (\omega).
}
inside the system. The CF functions and their conjugates are biorthogonal and
can be chosen to satisfy the condition
\eqn{
  \int_{\mc I} d \mbf r\, \bs \phi_k^* (\mbf r, \omega) \cdot \bs \psi_{k'}
  (\mbf r, \omega) = \delta_{kk'},
\label{biorth}
}
where the integration is over the interior~$\mc I$.

A Fourier component $\mbf E_{\omega} (\mbf r)$ of the internal field can be
expanded in the CF modes~as
\aln{
  &\mbf E_{\omega} (\mbf r) = \epsilon^{-1/2} (\mbf r) \sum_k a_k (\omega)\,
  {\bs \psi}_k (\mbf r, \omega),
\label{normmod} \\
  &a_k (\omega) = \int_{\mc I} d \mbf r\, \sqrt{\epsilon (\mbf r)}\, \bs
  \phi_k^* (\mbf r, \omega) \cdot \mbf E_{\omega} (\mbf r).
\label{normamp}
}
When continued to the exterior, this expansion yields a wave at the
frequency~$\omega$ propagating in the free space away from the system. In a
stationary lasing regime $\omega$ becomes subjected to an equation, which has a
discrete set of solutions corresponding to the frequencies of lasing modes.

\section{Semiclassical laser equations
\label{sec:sle}}

In the semiclassical theory of lasers~\cite{sarg74,hake84} the fields are
described classically at the level of Maxwell equations and the active medium
is treated by quantum mechanics. To this end, the wave equation~\eq{weqpas}
is written with a source term, the polarization $\mbf P (\mbf r, t)$ of the gain
medium:
\eqn{
  \epsilon (\mbf r)\, \frac {\pd^2} {\pd t^2} \mbf E + \nabla \times \l(\nabla
  \times \mbf E \r) = - 4 \pi \frac {\pd^2} {\pd t^2} \mbf P (\mbf r, t).
\label{weqact}
}
In the simplest model, the active medium is a collection of homogeneously
broadened two-level atoms. Their state is fully described by~$\mbf P (\mbf r,
t)$ and the population-inversion density~$\Delta n (\mbf r, t)$ (difference
between populations of the upper and lower levels per unit volume). These
functions satisfy the equations of motion~\cite{hake84}
\aln{
  &\l(\frac {\pd^2} {\pd t^2} + 2 \gamma_\perp \frac \pd {\pd t} + \nu^2 \r)
  \mbf P =  -2 \nu \frac {d^2} \hbar \mbf E (\mbf r, t)\, \Delta n (\mbf r, t),
\label{poleq} \\
  &\frac \pd {\pd t} \Delta n - \gamma_\parallel [\Delta n_0 (\mbf r, t) -
  \Delta n] = \frac 2 {\hbar \nu} \mbf E (\mbf r, t) \cdot \frac \pd {\pd t}
  \mbf P (\mbf r, t).
\label{inveq}
}
where $d$~is the magnitude of the atomic dipole matrix element, $\nu$~is the
atomic transition frequency, and $\gamma_\perp$ ($\gamma_\parallel$)~is the
polarization (population-inversion) decay rate. If the right-hand side of
Eq.~\eq{inveq} vanishes, $\Delta n$~relaxes to the unsaturated population
inversion~$\Delta n_0 (\mbf r, t)$, which is a measure of the pump strength.
The coupled equations~\eq{weqact}-\eq{inveq} determine, in principle, electric
field in the system, if $\Delta n_0 (\mbf r, t)$ is given.

It is convenient, at this stage, to rewrite the equations of motion in the
frequency representation. The Fourier-transformed variables are given in the
form
\aln{
  &\mbf E (\mbf r, t) = \frac 1 \pi \text{Re} \int_0^\infty d \omega\, \mbf
  E_\omega (\mbf r)\,  e^{-i \omega t}, \\
  &\mbf P (\mbf r, t) = \frac 1 \pi \text{Re} \int_0^\infty d \omega\, \mbf
  P_\omega (\mbf r)\,  e^{-i \omega t}, \\
  &\Delta n (\mbf r, t) = \frac 1 {2 \pi} \int_{-\infty}^\infty d \omega
  \Delta n_\omega (\mbf r)\, e^{-i \omega t}
}
that facilitates application of the rotating-wave approximation.
Additionally, we assume that $\mbf E_\omega$ and $\mbf P_\omega$ are negligible
outside of a small vicinity of~$\nu$. Then the frequency squared can be
approximated as $\omega^2 = (\nu + \omega - \nu)^2 \approx \nu^2 + 2 \nu
(\omega - \nu)$ and the lasing equations become
effectively the first-order differential equations in time,
\aln{
  &- \epsilon (\mbf r)\, (-\nu^2 + 2 \nu \omega)\, \mbf E_\omega + \nabla
  \times (\nabla \times \mbf E_\omega) = 4 \pi \nu^2 \mbf P_\omega,
\label{weqactft}\\
  &[-i (\omega - \nu) + \gamma_\perp ] \mbf P_\omega = - i \frac
  {d^2} {2 \pi \hbar} \int_0^\infty d \omega'\, \mbf E_{\omega'} \Delta
  n_{\omega - \omega'},
\label{poleqft}\\
  &(-i \omega + \gamma_\parallel) \Delta n_\omega =  2 \pi \gamma_\parallel
  \Delta n_0 (\mbf r)\, \delta (\omega) \notag \\
  &\quad - \frac i {\pi \hbar} \int_0^\infty
  d \omega' (\mbf E_{\omega' - \omega}^* \cdot \mbf P_{\omega'} - \mbf
  E_{\omega' + \omega} \cdot \mbf P_{\omega'}^*),
\label{inveqft}
}
where the pump $\Delta n_0 (\mbf r)$ is assumed constant in time.

\section{All-order perturbation theory}
\label{sec:all}

\subsection{Expansions for polarization and population inversion}

Equations \eq{weqactft}-\eq{inveqft} can be reduced to an equation for the
electric field alone using perturbation theory in the field amplitude. In
particular, one needs to construct an expansion of $\mbf P_\omega (\mbf r)$ in
the (odd) powers of the field using Eqs.~\eq{poleqft} and~\eq{inveqft}. Then
this expansion is substituted in Eq.~\eq{weqactft} producing the required
equation for the field. In the conventional laser theory~\cite{sarg74,hake84}
$\mbf P_\omega (\mbf r)$ is expanded up to the third order in~$\mbf E_\omega
(\mbf r)$, which yields the saturation terms in the rate equations. In this
paper we will carry out the expansion up to an arbitrary order in the field's
amplitude and use diagrammatic method to sort out the respective terms.

We begin by neglecting the quadratic terms in Eq.~\eq{inveqft} which gives
the zero-order expression for the population inversion: $\Delta n_\omega^{(0)} =
2 \pi \Delta n_0 (\mbf r)\, \delta (\omega)$. Substituting this expression in
the polarization Eq.~\eq{poleqft}, we obtain the first-order term for
polarization $\mbf P_\omega^{(1)} = - i (d^2 / \hbar \gamma_\perp)\, D(\omega)\,
\Delta n_0 (\mbf r)\, \mbf E_\omega$, where
\eqn{
  D(\omega) \equiv \l[1 - i \frac {\omega - \nu} {\gamma_\perp} \r]^{-1}.
}
This expression is substituted back in Eq.~\eq{inveqft}, from which the
second-order correction to the inversion $\Delta n_\omega^{(2)}$ is determined.
This iteration procedure yields the perturbation series for polarization and
population inversion,
\aln{
  &\mbf P_\omega (\mbf r) = \sum_{q \text{ odd}} \mbf P_\omega^{(q)} (\mbf r),
\label{Psum}  \\
  &\Delta n_\omega (\mbf r) = \sum_{q \text{ odd}} \Delta n_\omega^{(q - 1)}
  (\mbf r),
}
in odd and even powers of the electric field, respectively. Henceforth we
restrict the calculations to the case of scalar field. The general terms
of these expansions are derived by induction in Appendix~\ref{deriv}. The
resulting expressions are
\begin{widetext}
\aln{
  P_\omega^{(q)} =\, & 2 i \hbar \gamma_\parallel  \frac {A^{(q + 1)/2}}
  {\pi^{q - 1}} \Delta n_0 (\mbf r)\, D(\omega) \int d \omega_1 \cdots d
  \omega_{q - 1} E_{\omega_1} E_{\omega_2}^* \cdots E_{\omega_{q - 2}}
  E_{\omega_{q - 1}}^* E_{\omega - \omega_1 + \omega_2 - \cdots - \omega_{q -
  2} + \omega_{q - 1}} \notag \\
  &\times D_\parallel (\omega_1 - \omega_2)\, D_\parallel (\omega_1 - \omega_2
  + \omega_3 - \omega_4) \cdots D_\parallel (\omega_1 - \omega_2 + \cdots +
  \omega_{q - 2} - \omega_{q - 1}) \notag \\
  &\times [D(\omega_1) + D^*(\omega_2)] [D(\omega_1 - \omega_2 + \omega_3) +
  D^*(\omega_2 - \omega_1 + \omega_4)] \cdots \notag \\
  &\times [D(\omega_1 - \omega_2 + \omega_3 - \cdots - \omega_{q - 3} +
  \omega_{q - 2}) + D^*(\omega_2 - \omega_1 + \omega_4 - \cdots - \omega_{q -
  4} + \omega_{q - 1})],
\label{Pq}
}
\aln{
  \Delta n_\omega^{(q + 1)} =\, & 2 \frac {A^{(q + 1)/2}}{\pi^q} \Delta n_0
  (\mbf r)\, D_\parallel (\omega) \int d \omega_1 \cdots d \omega_q\,
  E_{\omega_1} E_{\omega_2}^* \cdots E_{\omega_{q - 1}}^* E_{\omega_q}
  E_{-\omega + \omega_1 - \omega_2 + \cdots - \omega_{q - 1} + \omega_q}^*
  \notag \\
  &\times D_\parallel (\omega_1 - \omega_2)\, D_\parallel (\omega_1 - \omega_2
  + \omega_3 - \omega_4) \cdots D_\parallel (\omega_1 - \omega_2 + \cdots +
  \omega_{q - 2} - \omega_{q - 1}) \notag \\
  &\times [D(\omega_1) + D^*(\omega_2)] [D(\omega_1 - \omega_2 +
  \omega_3) + D^*(\omega_2 - \omega_1 + \omega_4)] \cdots \notag \\
  &\times [D(\omega_1 - \omega_2 + \omega_3 - \cdots - \omega_{q - 3} +
  \omega_{q - 2})  +  D^*(\omega_2 - \omega_1 + \omega_4 - \cdots - \omega_{q -
  4} + \omega_{q - 1})] \notag \\
  &\times D(\omega_1 - \omega_2 + \omega_3 - \cdots - \omega_{q - 1} +
  \omega_q) + \text{c.c.}(\omega \to -\omega),
\label{nq}
}
\end{widetext}
where $q$ is odd. The notation $\text{c.c.}(\omega \to -\omega)$ stands for the
first part of the equation to which complex conjugation accompanied by change of
the sign of~$\omega$ was applied. We also introduced the following definitions
\aln{
  &A \equiv - \frac {d^2} {2\hbar^2 \gamma_\perp \gamma_\parallel}, \\
  &D_\parallel (\omega) \equiv \l[1 - i \frac \omega {\gamma_\parallel}
  \r]^{-1}.
}
The above expressions describe nonlinear ($q \geq 3$) corrections to
polarization and population inversion, which are used below to obtain equations
for the field amplitudes.

\subsection{Equations for electric field}

Equation for the amplitudes~\eq{normamp} of normal modes,
\eqn{
  [\Omega_k (\omega) - \omega] a_k (\omega) = 2 \pi \nu \int_{\mc I} d \mbf r\,
  \epsilon^{-1/2} (\mbf r)\, \phi_k^* (\mbf r, \omega)\, P_\omega (\mbf r),
\label{eomnorm}
}
follows from Eq.~\eq{weqactft} and biorthogonality condition~\eq{biorth}. The
right-hand side is a sum of the infinite number of nonlinear corrections to the
polarization and is, therefore, a functional of \emph{all} amplitudes~$a_k
(\omega)$. By definition, a lasing mode is such a solution which, in the limit
$t \to \infty$, approaches a purely harmonic form, $\propto \exp(-i\omega_k t)$.
Here $\omega_k$ is some frequency, which is determined self-consistently from
Eq.~(\ref{eomnorm}). In the frequency domain these solutions are characterized
by delta-functional frequency dependence $\propto \delta (\omega - \omega_k)$.

Generally speaking, lasing modes are different from the quasimodes of the
passive system. Indeed, an attempt to search for the solutions of
Eq.~\eq{eomnorm} in the form $a_k (\omega) = a_k\, \delta (\omega - \omega_k)$
leads, after the frequency integration~\eq{Pq}, to the equation
\aln{
  [\Omega_k &(\omega_k) - \omega_k]\, a_k\, \delta (\omega - \omega_k) \notag
  \\
  = &\sum_{q \text{ odd}}\: \sum_{ k_1, \ldots, k_q}  W_{k, k_1, \ldots,
  k_q}^{(q)} a_{k_1} a_{k_2}^* \cdots a_{k_{q-1}}^*a_{k_q}\notag \\
  &\times  \delta (\omega  - \omega_{k_1} + \omega_{k_2} - \cdots +
  \omega_{k_{q-1}} - \omega_{k_q}),\label{eqomnonres}
}
where the coefficients $W_{k, k_1, \ldots, k_q}^{(q)}$ are functions of the
frequencies $\omega_k, \omega_{k_1}, \ldots, \omega_{k_q}$. Clearly, most of
the terms on the right-hand side contain the delta functions that are different
from $\delta (\omega - \omega_k)$ on the left-hand side. In the time domain,
these terms would introduce oscillations with frequencies different
from~$\omega_k$. This means that, in general, no stationary lasing solutions
exist unless, for some reasons, the terms oscillating at the ``beat''
frequencies are small and can be neglected. Usually the selection of the slowly
changing contributions to Eq.~\eq{eqomnonres} is done by leaving only terms
with pairwise coinciding indices~$k_i$, so that the respective frequency
differences cancel out (more on this procedure can be found below).
However, since the number of $\omega_{k_i}$ in this equation is odd, one will
always remain with the expression $\delta(\omega - \omega_{k^\prime})$, where
$k^\prime$ is the index of one of the remaining uncanceled frequencies.

This problem can only be resolved by requiring that, in a given lasing mode,
$l$, each contribution $a_{lk}$ of the quasimode~$k$ oscillates at the same
frequency, and the general solution for $a_k$ takes the form~of
\eqn{
  a_k (\omega) = \pi \sum_l a_{lk}\, \delta(\omega-\omega_l).
\label{modexp}
}
Amplitudes  $a_{lk}$ can be shown to obey the equation
\eqn{
  [\Omega_k (\omega_l) - \omega_l]\, a_{lk}  =
  \sum_{k^\prime} V_{kk^\prime}^{(l)} \, a_{lk^\prime},
\label{EigenValue}
}
where the coefficients~$V_{kk'}^{(l)}$ depend on all frequencies and
amplitudes. An explicit expression for $V_{kk'}^{(l)}$ is given below. One can
see from this equation that the lasing modes are those combinations of the
quasimodes that diagonalize the matrix $V_{kk'}^{(l)}$, while lasing
frequencies are real eigenvalues of this matrix~\cite{deyc05a,deyc05b}.

One has to realize, though, that Eq.~(\ref{EigenValue}) is not an ordinary
eigenvalue problem, since the matrix $V_{kk'}^{(l)}$ itself depends on the
amplitudes~$a_{lk}$. Unlike linear eigenvalue problems, that determine
frequencies for which nonzero solutions for the amplitudes can exist, solving
Eq.~(\ref{EigenValue}) one must be able to find the frequencies, as well as the
field amplitudes. This is possible, because the requirement that the respective
eigenfrequencies must be real provides an additional constraint on solutions of
Eq.~(\ref{EigenValue})~\cite{ture08,ture09}.

If matrix $V_{kk'}^{(l)}$ is calculated in the constant-inversion approximation
(see below), Eq.~(\ref{EigenValue}) reproduces the main result of
Ref.~\cite{ture07}. However, while the derivation of this equation in
Ref.~\cite{ture07} is only valid in the strictly stationary limit, the
arguments presented here can be extended to a nonstationary case. Indeed, we
can repeat the above arguments for a weakly nonstationary situation, requiring
that the amplitudes $a_{lk}$ be slowly changing functions of time. Formally, we
replace the mode expansion~\eq{modexp} with $a_k (\omega) = \sum_l a_{lk}
(\omega - \omega_l)$, where $a_{lk} (\omega - \omega_l)$ is assumed to be
sharply peaked at $\omega = \omega_l$. In this case we can transform
Eq.~\eq{eomnorm} to the time domain by expanding $\Omega_k (\omega)$ as
$\Omega_k \approx \Omega_k (\omega_l) + (\omega-\omega_l) \Omega_k^\prime
(\omega_l)$, where $\Omega_k^\prime (\omega)$ is the derivative of $\Omega_k$
with respect to the spectral parameter~$\omega$. It was found in
Ref.~\cite{shuv09} that, at least in the case of modes of a disk resonator,
this derivative is not small and must be taken into account. In nonlinear terms
we simply replace $a_{lk} (\omega - \omega_l) \to \pi a_{lk} (t)\,
\delta(\omega - \omega_l)$, which amounts to neglect of time derivatives of the
nonlinear corrections. The resulting equation~is
\mul{
  \l\{-i \l[1 - \Omega_k^\prime\r (\omega_l)]\frac d {dt} +  [\Omega_k
  (\omega_l) - \omega_l] \r\} a_{lk} (t) \\
  = \sum_{k'} V_{kk'}^{(l)} (t)\, a_{lk'}(t),
\label{eomlas}
}
which, in the stationary limit, coincides with Eq.~(\ref{EigenValue}). Note
that $V_{kk'}^{(l)}$ is now a slowly varying function of time via its
dependence on the amplitudes~$a_{l''k''} (t)$.  The correction due to
$\Omega_k^\prime$ is a new term, which has not been discussed in any of the
previous treatments of lasing dynamics. While it does not affect the
steady-state solutions, it might change their stability, and is, therefore,
important for strongly open cavities. More detailed study of its role is
outside of the scope of this paper and will be presented elsewhere.

The polarization matrix $V_{kk'}^{(l)}$ in the slowly varying amplitude
approximation can be presented~as
\eqn{
   V_{kk'}^{(l)} (t) = 2 \pi \nu\int_{\mc I} d \mbf r\, \epsilon^{-1} (\mbf r)\,
  \phi_k^* (\mbf r, \omega_l)\, \psi_{k'} (\mbf r,\omega_l)\, \eta_l (\mbf{r},
  t),
\label{Vkk}
}
where we introduced the nonlinear susceptibility $\eta_l (\mbf{r}, t) \equiv
P_l (\mbf{r}, t)/E_l (\mbf{r}, t)$ defined as the ratio of the slowly varying
polarization amplitude $P_l (\mbf{r}, t)$ and the field $E_l (\mbf{r}, t)$ in
the mode~$l$. The expression for the susceptibility is found from perturbation
expansion for polarization $P(\mbf r, t)$, Eqs.~\eq{Psum} and~\eq{Pq}, and is
given~by
\begin{widetext}
\aln{
  \eta_l (\mbf{r}, t) = \, & 2i  \hbar \gamma_\parallel D(\omega_l) \Delta n_0
  (\mbf r)  \sum_{q \text{ odd}} A^{\frac {q + 1} 2}\sideset{}{^{\text r}}
  \sum_{l_1, \ldots, l_q} |E_{l_2} (\mbf r, t)|^2\, |E_{l_4} (\mbf r, t)|^2
  \cdots |E_{l_{q - 1}} (\mbf r, t)|^2 \notag \\
  &\times D_\parallel (\omega_{l_1} - \omega_{l_2})\, D_\parallel (\omega_{l_1}
  - \omega_{l_2} + \omega_{l_3} - \omega_{l_4}) \cdots D_\parallel
  (\omega_{l_1} - \omega_{l_2} + \cdots + \omega_{l_{q - 2}} - \omega_{l_{q -
  1}})\notag \\
  &\times [D(\omega_{l_1}) + D^*(\omega_{l_2})] [D(\omega_{l_1} - \omega_{l_2}
  + \omega_{l_3}) + D^*(\omega_{l_2} - \omega_{l_1} + \omega_{l_4})] \cdots
  \notag \\
  &\times [D(\omega_{l_1} - \omega_{l_2} + \omega_{l_3} - \cdots - \omega_{l_{q
  - 3}} + \omega_{l_{q - 2}}) + D^*(\omega_{l_2} - \omega_{l_1} + \omega_{l_4}
  - \cdots - \omega_{l_{q - 4}} + \omega_{l_{q - 1}})].
\label{Pl}}
\end{widetext}
Here the order of nonlinearity $q$, introduced in Eq.~\eq{Psum}, determines the
number of different indices $l_i$, which take values from $1$ to~$N_{\text m}$,
where $N_{\text m}$ is the number of lasing modes. The superscript~``r'' at the
sum symbol specifies that the possible values of the indices are restricted by
the resonance condition
\eqn{
  \omega_{l_1} - \omega_{l_2} + \omega_{l_3} - \cdots - \omega_{l_{q - 1}} +
  \omega_{l_q} - \omega_l = 0
\label{res}
}
which ensures cancelation of fast oscillating terms. In the absence of
accidental degeneracies, this condition implies that each of the indices $l_1,
l_3, \ldots, l_q$ must be equal to one of the indices $l_2, l_4, \ldots, l_{q -
1}, l$. This leads, in particular, to the appearance of absolute squares of the
field in the first line of Eq.~\eq{Pl}. Moreover, the index $l_q$ effectively
drops out of the equation, since the amplitude $E_{l_q}$ must be equal to some
other amplitude~$E_{l_i}$. It is assumed that the slowly varying field
amplitudes are expressed in terms of quasimode components $a_{lk} (t)$ of the
respective $l$th lasing mode using
\eqn{
  E_l (\mbf r,t) = \epsilon^{-1/2} (\mbf r) \sum_k a_{lk}(t)\, \psi_k (\mbf r,
  \omega_l).
\label{Eexpst}
}
Substituting Eqs.~\eq{Pl} and~\eq{Vkk} in Eq.~\eq{eomlas} one
obtains a closed system of dynamic equations for $a_{lk}$ valid to all orders in
the field amplitude.

One of the fundamental difficulties of the theory of lasers is that the
number~$N_{\text m}$ of lasing modes is a priori unknown and depends on the
strength of the pumping and the spatial distribution of the electric field in
the cavity. In Ref.~\cite{ture07} this value is determined by the number of
possible solutions of Eq.~\eq{EigenValue} with real frequencies at a given
pumping strength. This approach, however, does not take into account stability
of the found solutions, which can only be determined by considering the
time-dependent Eq.~\eq{eomlas}. Using this equation one could start by assuming
that $N_{\text m}$ is equal to the size of the basis of quasimodes, $N_{\text
b}$, and study their time evolution. Those $E_l$ which do not correspond to
real lasing solutions at given pumping would decay to zero, and the number of
lasing modes would be determined a posteriori without the need for a prior
knowledge of~$N_{\text m}$. This approach is not free of difficulties either,
because of possible multistable behavior and hysteresis. Analysis of these
issues, however, is beyond the scope of this paper.

\section{Diagrammatic technique
\label{sec:dt}}

\subsection{Diagrammatic representation of the perturbation series}

In this subsection we present a diagrammatic technique developed to classify
different nonlinear terms in Eq.~\eq{Pl}. It should be noted, however, that our
diagrams, unlike diagrams of the field or many-particle theory, do not provide
one-to-one correspondence between different terms of Eq.~\eq{Pl} and elements
of the diagrams. The role of the diagrams here is more limited: we use them to
classify different pairing possibilities for the lasing mode indices $l_1, l_2,
\ldots, l_q, l$ in the perturbation series~\eq{Pl}. Nevertheless, as it is
shown below, this technique allows for classification and partial summation of
the classes of the terms in a manner very similar to traditional diagrammatic
methods. Unlike pairing of vertices in traditional diagrammatic techniques,
which reflects Vick's theorem for creation-annihilation operators or Gaussian
statistics of respective random processes, the pairing procedure in the
situation under consideration hinges upon the condition expressed by
Eq.~\eq{res}. The resonance condition guarantees the absence of the fast
oscillating terms, and, hence, the validity of the slowly changing amplitude
approximation.

\begin{figure}
  \includegraphics[width = .35 \linewidth]{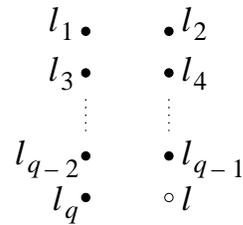}
  \caption{Labeling of vertices in a diagram of order $q = 1, 3, \ldots$}
\label{labels_fig}
\end{figure}

\begin{figure}
  \includegraphics[width = .55 \linewidth]{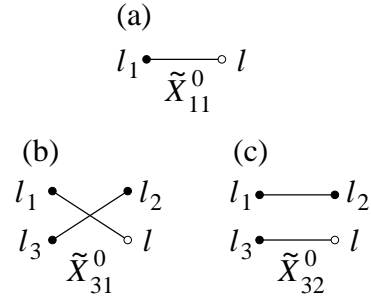}
  \caption{First-order diagram (a) and third-order diagrams~(b,~c).}
\label{diag_1_3_fig}
\end{figure}

\begin{figure}
  \includegraphics[width = .7 \linewidth]{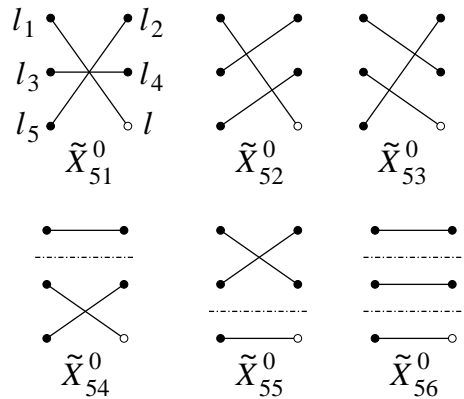}
  \caption{Fifth-order diagrams. The dash-dotted lines are the cuts that split
  disconnected diagrams into connected diagrams.}
\label{diag_5_cut_fig}
\end{figure}

To construct a diagram $\wtilde X_{qj}^0$ of order $q = 1, 3, \ldots$, we place
$q + 1$ vertices in two columns as shown in Fig.~\ref{labels_fig}. The left
vertices are labeled $l_1, l_3, \ldots, l_q$ and the right vertices are labeled
$l_2, l_4, \ldots, l_{q - 1}, l$. The vertex $l$ is different from the other
vertices, because there is no summation over the index $l$ in Eq.~\eq{Pl}. After
that, each vertex on the left is connected with exactly one vertex on the right.
The index $j = 1, \ldots, \wtilde N_q$ labels all distinct connection
possibilities in an arbitrary order. To obtain all diagrams of order~$q$, we
can, first, connect the vertices by $(q + 1)/2$ horizontal links and then
reshuffle the vertices, say, on the left without cutting the links. Thus, the
number of possible diagrams of order~$q$ is the number of permutations $\wtilde
N_q = [(q + 1)/2]!$. Diagrams for $q = 1, 3, 5$ are shown in
Figs.~\ref{diag_1_3_fig} and~\ref{diag_5_cut_fig}.

Each diagram specifies a particular contribution to the series~\eq{Pl}. The
latter will be written in the form
\aln{
  &\eta_l (\mbf r, t) = 2 i \hbar \gamma_\parallel \Delta n_0 (\mbf r)\,
  D(\omega_l) \mc X_l,
\label{Pldiag}\\
  &\mc X_l \equiv \sum_{q \text{ odd}}A^{\frac {q + 1} 2}\, \sum_{j =
  1}^{\wtilde N_q} \wtilde X_{qj}^0.
\label{diagser}
}
where each $\wtilde  X_{qj}^0$ represents a partial sum in $\sideset{}{^{\text
r}} \sum(\cdots)$, Eq.~\eq{Pl}, in which pairs of indices are chosen to be
equal to each other according to the links connecting respective vertices in
the diagram. For example, the first three diagrams correspond to the following
expressions:
\aln{
  \wtilde X_{11}^0 = &1,
\label{X110}\\
  \wtilde X_{31}^0 = &\sum_{l_2 \ne l} |E_{l_2} (\mbf r, t)|^2 D_\parallel
  (\omega_l - \omega_{l_2})  [D(\omega_l) + D^*(\omega_{l_2})],
\label{X310}\\
  \wtilde X_{32}^0 = &\sum_{l_2} |E_{l_2} (\mbf r, t)|^2\, 2 \text{Re}
  [D(\omega_{l_2})].
}
The restriction $l_2 \ne l$ in the diagram $\wtilde X_{31}^0$ excludes the term
with $l_1 = l_2 = l_3 = l$, which enters~$\wtilde X_{32}^0$. In general, the
terms with more than two indices equal belong to the diagram in which the links
connecting these indices do not cross each other. Another example are the
fifth-order terms with $l_2 = l_3 = l_4 = l_5 \ne l$, which enter $\wtilde
X_{52}^0$, but not~$\wtilde X_{51}^0$. Expression for arbitrary $\wtilde
X_{qj}^0$ is given in Appendix~\ref{genexpr}.

\subsection{Resummation of the diagrams}

A diagram is called connected if it cannot be cut by a horizontal line without
cutting a link. For instance, the diagrams $\wtilde X_{31}^0$, $\wtilde
X_{51}^0$, $\wtilde X_{52}^0$, and $\wtilde X_{53}^0$ are connected, while the
diagrams $\wtilde X_{32}^0$, $\wtilde X_{54}^0$, $\wtilde X_{55}^0$, and
$\wtilde X_{56}^0$ are disconnected. To simplify the notation, we ordered all
connected diagrams before the disconnected diagrams for given~$q$. We will label
connected diagrams as $X_{qj}^0$ with
\eqn{
  X_{qj}^0 = \wtilde X_{qj}^0, \quad j = 1, \ldots, N_q,
\label{conndiag}
}
where $N_q$ ($< \wtilde N_q$) is the number of connected diagrams.  The
horizontal cuts separate disconnected diagrams into one connected diagram
containing the vertex~$l$ (denoted by an unfilled dot in the graphic
representation) and several connected subdiagrams without such vertex. The
latter subdiagrams will be denoted as $X_{qj}$, $j = 1, \ldots, N_q$, where $q +
1$ is the number of vertices in the subdiagram. In place of the vertex~$l$ these
diagrams have a vertex with the index $l_{q + 1}$ which runs over all lasing
modes, as the other indices~$l_j$. For example, the diagram $\wtilde X_{55}^0$
consists of $X_{11}^0$ and $X_{31}$ and the diagram $\wtilde X_{56}^0$ consists
of $X_{11}^0$ and two diagrams $X_{11}$, where
\aln{
  &X_{11} = \sum_{l_2} |E_{l_2} (\mbf r, t)|^2\, 2 \text{Re} [D(\omega_{l_2})],
  \\
  &X_{31} = \sum_{\mbox{\scriptsize
  $\begin{array}{c}
    l_2, l_4 \\
    l_2 \ne l_4
  \end{array}$
  }} |E_{l_2} (\mbf r, t)|^2
  |E_{l_4} (\mbf r, t)|^2 \notag \\
  &\phantom{X_{31} =}\times D_\parallel (\omega_{l_4} - \omega_{l_2})
  [D(\omega_{l_4}) + D^*(\omega_{l_2})]^2.
\label{X31}
}
A general expression for $X_{qj}$ is given in Appendix~\ref{genexpr}. Note that
$X_{qj}$ is of the order $q + 1$ in the electric field. Connected diagrams
contain $(q - 1)/2$ nontrivial factors $D_\parallel \ne 1$.

Our diagrammatic technique possesses the basic property that
disconnected diagrams are given by products of their connected parts. Thus, we
can write for our examples
\aln{
  &\wtilde X_{55}^0 = X_{11}^0\, X_{31}, \\
  &\wtilde X_{56}^0 = X_{11}^0\, (X_{11})^2.
}
The multiplicativity is due to the fact that the resonance condition of the
type~\eq{res} is fulfilled for each connected subdiagram (see also
Appendix~\ref{genexpr}). The multiplicativity property allows to express the
series~\eq{diagser} in terms of the connected diagrams~as
\aln{
  \mc X_l &= \l(\sum_{q \text{ odd}}\, \sum_{j = 1}^{N_q} X_{qj}^0 \r)\,
  \sum_{m = 0}^\infty \l(\sum_{q \text{ odd}}\, \sum_{j = 1}^{N_q} X_{qj} \r)^m
  \notag \\
  &= \frac {\sum_{q \text{ odd}}\, \sum_{j = 1}^{N_q} X_{qj}^0}
  {1 - \sum_{q \text{ odd}}\, \sum_{j = 1}^{N_q} X_{qj}}.
\label{resum}
}
This resummation formula is the main result of our paper.

\section{Limiting cases and discussion
\label{sec:lim}}

To make the meaning of Eq.~\eq{resum} more transparent we apply it in several
well-known special cases. We start with the linear approximation in
Sec.~\ref{lin} and consider the effect of gain-induced coupling of passive
modes~\cite{deyc05a,deyc05b}. In Sec.~\ref{third} we reproduce  semiclassical
equations of the standard third-order laser theory~\cite{sarg74,hake84}. In
Sec.~\ref{const} we discuss all-order nonlinear theory in the approximation of
constant population inversion~\cite{ture06,ture07,ture08,ture09} and derive,
using our theory, the first diagrammatic correction to~it.

\subsection{Linear gain-induced mode coupling
\label{lin}}

In the linear approximation to the polarization $P_l = \eta_l E_l$ [Eq.~\eq{Pl}]
only the lowest diagram~$X_{11}^0$ contributes to~$\mc X_l$.
Equation~\eq{eomlas}, where the matrix $V_{kk'}^{(l)}$ is calculated using
Eqs.~\eq{Vkk}, \eq{Pldiag}, and~\eq{X110}, yields the following equations for
the slowly-varying amplitudes:
\eqn{
  \sum_{k'} \l\{ \delta_{kk'} \frac d {dt} + i \l[ \wbar \Omega_{kk'} (\omega_l)
  - \delta_{kk'} \omega_l \r] \r\} a_{lk'} = 0,
\label{lineq}
}
where the $\Omega'_k$ term is henceforth neglected.  The frequency matrix
\eqn{
  \wbar \Omega_{kk'} (\omega) = \Omega_k  (\omega)\, \delta_{kk'} -
  V_{kk'} (\omega),
}
is modified by the linear gain term,
\mul{
  V_{kk'} (\omega) =  \\
  - 2 i \pi \nu \frac {d^2} {\hbar \gamma_\perp}
  D(\omega)\int_{\mc I} d \mbf r \frac {\Delta n_0 (\mbf r)} {\epsilon (\mbf
  r)}\, \phi_k^* (\mbf r, \omega)\, \psi_k (\mbf r, \omega),
}
proportional to the overlap integral. Clearly, the matrices $V_{kk'}$
and~$\wbar \Omega_{kk'}$ are nondiagonal if the pump or dielectric constant are
not homogeneous. In this case the biorthogonal quasimodes of the system $\wbar
\psi_k$ and~$\wbar \phi_k$ are no longer $\psi_k$ and~$\phi_k$, but are
determined by the right and left eigenvectors~$a_{lk}^{\text{(r,l)}}$ of~$\wbar
\Omega_{kk'}$ according~to
\aln{
  &\wbar \psi_l (\mbf r, \omega) = \sum_{k'} a_{lk'}^{(\text r)}\, \psi_{k'}
  (\mbf r, \omega), \\
  &\wbar \phi_l (\mbf r, \omega)  = \sum_{k'} a_{lk'}^{(\text l)}\, \phi_{k'}
  (\mbf r, \omega).
}
The right eigenvectors $a_{lk'}^{(\text r)}$ are normal modes of
Eq.~\eq{lineq} whose amplitudes~$\wbar a_l (t)$ obey the equation
\eqn{
  \dot {\wbar a}_l + i \l[ \wbar \Omega_l (\omega_l) - \omega_l \r] \wbar a_l
  = 0,
\label{newmod}
}
where $\wbar \Omega_l (\omega)$ are eigenvalues of~$\wbar \Omega_{kk'}
(\omega)$. We recall that electric field in the mode~$l$ has a time dependence
$\wbar a_l (t) \exp (- i \omega_l t)$. Thus, the lasing frequency~$\omega_l$ is
determined from the requirement
\eqn{
  \text{Re} [\wbar \Omega_l (\omega_l)] = \omega_l
}
and the threshold condition for this mode~is
\eqn{
  \text{Im} [\wbar \Omega_l (\omega_l)] = 0.
}
As follows from Eq.~\eq{newmod}, the mode amplitudes diverge exponentially
above the threshold. Hence, applicability of the linear approximation is
limited to the pump strength below or at the threshold. However, the basis of
normal modes can be used as a starting point in nonlinear theories.

\subsection{Third-order theory
\label{third}}

To obtain an approximation to $\mc X_l$ of the third-order in the field, we
keep the diagrams $X_{11}^0$ and $X_{31}^0$ in the numerator and the diagram
$X_{11}$ in the denominator of Eq.~\eq{resum} and expand the latter:
\eqn{
  \mc X_l \approx X_{11}^0 + X_{11}^0 X_{11} + X_{31}^0.
}
It is convenient to write lasing equations in the basis of quasimodes $\wbar
\psi_k$ and~$\wbar \phi_k$ that diagonalize the linear part. Due to nonlinear
effects, the lasing modes above the threshold,
\eqn{
  E_l (\mbf r, t) = \epsilon^{-1/2} (\mbf r) \sum_k \wbar a_{lk} (t)\, \wbar
  \psi_k (\mbf r, \omega_l),
\label{Elbar}
}
are, in general, linear combinations of individual quasimodes. Equation for the
amplitudes $\wbar a_{lk} (t)$ follows from Eq.~\eq{eomlas}, after taking into
account the results of linear theory (Sec.~\ref{lin}), and has the form:
\begin{widetext}
\aln{
  &\dot {\wbar a}_{lk} + i \l[ \wbar \Omega_k (\omega_l) - \omega_l \r] \wbar
  a_{lk} = - \frac {\pi \nu} {\hbar \gamma_\parallel} \l( \frac {d^2} {\hbar
  \gamma_\perp} \r)^2 D(\omega_l) \int_{\mc I} d \mbf r \frac {\Delta n_0 (\mbf
  r)} {\sqrt {\epsilon(\mbf r)}} \, \wbar \phi_k^* (\mbf r, \omega_l)\, E_l
  (\mbf r, t)  \notag \\
  &\times \sum_{l' = 1}^{N_{\text m}} |E_{l'} (\mbf r, t)|^2 \l\{ 2 \text{Re}
  [ D(\omega_{l'}) ] + (1
  - \delta_{ll'}) D_\parallel (\omega_l - \omega_{l'})\, [D(\omega_l) +
  D^*(\omega_{l'})] \r\}, \quad k = 1, \ldots, N_{\text b},
\label{leq3ord}
}
\end{widetext}
where $N_{\text b}$ is the size of the basis of quasimodes and $N_{\text m}$ is
the number of lasing modes. The lasing frequencies~$\omega_l$ and the mode
thresholds need to be determined from these $N_{\text b} \times N_{\text m}$
equations in the stationary regime $\dot {\wbar a}_{lk} = 0$ using, e.g., a
selfconsistent procedure described in Ref.~\cite{ture08}.

In some cases the standard assumption of a traditional lasing theory
that the lasing modes coincide with the quasimodes of the cavity remain valid.
In this case the total number of equations~\eq{leq3ord} is reduced to~$N_{\text
b}$ since the amplitudes are approximated as $\wbar a_{lk} (t) = \wbar a_l
(t)\, \delta_{lk}$.  Representing $\wbar a_l (t) = \sqrt{I_l} \exp (i \wbar
\varphi_l)$ and separating the real and imaginary parts in Eq.~\eq{leq3ord}, we
obtain $2N_{\text b}$ real equations for the intensities~$I_l$ and
phases~$\wbar \varphi_l$,
\aln{
  &\dot I_l - 2\text{Im} [\wbar \Omega_l (\omega_l)]\, I_l \notag \\
  &= - \frac {2 \pi \nu} {\hbar \gamma_\parallel} \l( \frac {d^2} {\hbar
  \gamma_\perp} \r)^2 I_l\, \text{Re} \l[D(\omega_l) \sum_{l'} B_{ll'}
  I_{l'}  \{ \cdots \} \r],
\label{rateI}\\
  &\dot{\wbar \varphi}_l + \text{Re} [\wbar \Omega_l (\omega_l)] - \omega_l
  \notag \\
  &= - \frac {\pi \nu} {\hbar \gamma_\parallel} \l( \frac {d^2} {\hbar
  \gamma_\perp} \r)^2 \text{Im} \l[D(\omega_l) \sum_{l'} B_{ll'}
  I_{l'}  \{ \cdots \} \r],
\label{rateom}
}
The terms enclosed in the braces are the same as those in Eq.~\eq{leq3ord}. We
defined overlap integrals for the quasimodes~as
\eqn{
  B_{ll'} = \int_{\mc I} d \mbf r \frac {\Delta n_0 (\mbf
  r)} {[\epsilon(\mbf r)]^2} \, \wbar \phi_l^* (\mbf r, \omega_l)\,
  \wbar \psi_l (\mbf r, \omega_l)\, |\wbar \psi_{l'}|^2.
}
The lasing frequencies $\omega_l$ are determined, together with the stationary
intensities, from Eqs.~\eq{rateI} and~\eq{rateom} in the stationary regime
$\dot I_l = 0$ and $\dot{\wbar \varphi}_l = 0$. The rate equations~\eq{rateI}
and frequency equations~\eq{rateom} generalize the standard third-order
semiclassical theory with the self- and cross-saturation
terms~\cite{sarg74,hake84} to systems with strong openness and arbitrary
distribution of refractive index.

\subsection{Constant-inversion approximation and corrections
\label{const}}

In many physical situations the population inversion $\Delta n (\mbf r,t)$ is
approximately constant in time. Oscillations of the population (population
pulsations) are responsible for the terms proportional to $D_\parallel (\omega_l
- \omega_{l'})$, $l \ne {l'}$, in the expansion of the nonlinear
susceptibility~\eq{Pl} and, hence, in the expressions for the diagrams.
Comparing the terms in the braces in Eq.~\eq{leq3ord}, we can conclude that the
population pulsations can be neglected if $D_\parallel (\omega_l - \omega_{l'})
\ll 1$, i.e., $|\omega_l - \omega_{l'}| \gg \gamma_\parallel$ for all lasing
frequencies $\omega_l \ne \omega_{l'}$.

In the approximation of constant inversion only the diagrams $X_{11}^0$ and
$X_{11}$, which do not contain the $D_\parallel$ functions, contribute to $\mc
X_l$~\eq{resum}. Equation~\eq{eomlas} for the coefficients of
expansion~\eq{Eexpst} of the electric field becomes
\mul{
  \dot a_{lk} + i [\Omega_k (\omega_l) - \omega_l] a_{lk}
  = 2 \pi \nu \frac {d^2} {\hbar \gamma_\perp} \int_{\mc I} d \mbf r
  \frac {\Delta n_0 (\mbf r)} {\sqrt {\epsilon(\mbf r)}}  \\
  \times \frac {\phi_k^*
  (\mbf r, \omega_l)\, E_l (\mbf r, t)} {1 + \frac {d^2} {2 \hbar^2
  \gamma_\perp \gamma_\parallel} \sum_{l'} |E_{l'} (\mbf r, t)|^2 2 \text{Re}
  [ D(\omega_{l'}) ]}.
\label{constpop}
}
In contrast to Eq.~\eq{leq3ord}, the linear contribution here is not
diagonalized and is contained in the right-hand side.

Equation~\eq{constpop} is valid in all orders in nonlinearity if the population
inversion is constant. With the help of Eq.~\eq{resum} it is straightforward to
write out the corrections due to the population pulsations. The terms of the
first order in $D_\parallel \ll 1$ are contained in the diagrams $X_{31}^0$
and~$X_{31}$ [Eqs.~\eq{X310} and~\eq{X31}], so that $\mc X_l$ can be
approximated~as
\eqn{
  \mc X_l \approx \frac {X_{11}^0 + X_{31}^0} {1 - X_{11} - X_{31}}.
}
These corrections modify both the numerator and denominator of
Eq.~\eq{constpop}, which can now be presented~as
\mul{
  \dot a_{lk} + i [\Omega_k (\omega_l) - \omega_l] a_{lk}
  = 2 \pi \nu \frac {d^2} {\hbar \gamma_\perp} \int_{\mc I} d \mbf r
  \frac {\Delta n_0 (\mbf r)} {\sqrt {\epsilon(\mbf r)}}  \\
  \times \frac {\phi_k^* (\mbf r, \omega_l)\, E_l (\mbf r,
  t)\l(1-\Upsilon_{\text n} \r)} {1 + \frac {d^2} {2 \hbar^2 \gamma_\perp
  \gamma_\parallel} \sum_{l'} |E_{l'} (\mbf r, t)|^2 2 \text{Re} [
  D(\omega_{l'}) ]\l(1-\Upsilon_{\text d} \r)}.
}
where
\aln{
  \Upsilon_{\text n} = \frac{d^2} {2\hbar^2 \gamma_\perp \gamma_\parallel}
  \sum_{l' \ne l} &|E_{l'} (\mbf r, t)|^2 D_\parallel (\omega_l - \omega_{l'})
   \notag\\
  &\times [D(\omega_l) + D^*(\omega_{l'})], \\
  \Upsilon_{\text d} = \frac {d^2} {2\hbar^2 \gamma_\perp \gamma_\parallel}
  \sum_{l''\ne l'} &|E_{l''} (\mbf r, t)|^2 D_\parallel (\omega_{l''} -
  \omega_{l'}) \notag \\
  &\times \frac{[D(\omega_{l''}) + D^*(\omega_{l'})]^2}{2 \text{Re}  [
  D(\omega_{l'}) ]}.
}
Taking into account that the electric field in the lasing modes has a typical
magnitude of $\hbar \sqrt{\gamma_\perp \gamma_\parallel}/d$, one can see that
deviations from the constant-inversion approximation is of the order of
$D_\parallel (\omega_l - \omega_{l'})$ [the term $D(\omega_l)$ is of the order
of unity since frequencies of lasing modes are concentrated within the width of
the gain profile]. The difference between lasing frequencies can be estimated
as $|\omega_l - \omega_{l'}| \approx \gamma_\perp/N_{\text m}$. Then the
condition $D_\parallel \ll 1$ can be expressed as
$N_{\text m}\gamma_\parallel / \gamma_\perp \ll 1$. Given that
$\gamma_\parallel$ is usually several orders of magnitude smaller than
$\gamma_\perp$, this condition is in most situations fulfilled. It was reported
in Ref.~\cite{ture09}, however, that nonlinear interaction between lasing modes
can push their frequencies toward each other making the intermode spectral
interval much smaller than the typical value given above. Such pairs of modes
can result in significant corrections to the constant-population approximation.
Adding to the expansion additional connected diagrams with up to $q + 1$
vertices, one can improve the constant-population approximation by constructing
lasing equations valid in the order $(q - 1)/2$ in~$D_\parallel$.

\section{Conclusions}

We presented a diagrammatic semiclassical laser theory valid in all orders of
electric field. The original perturbation series in the powers of the field can
be resummed in terms of a certain class of diagrams, the connected diagrams.
The resummation allows to construct a controlled expansion in the small
parameter $\gamma_\parallel / \gamma_\perp$, which is a measure of population
pulsations, while treating the nonlinearity exactly. Our lasing equations
generalize the all-order nonlinear equations in the constant-inversion
approximation and the third-order equations with population-pulsation terms.
The use of constant-flux quasimodes as basis functions makes it possible to
apply the theory to strongly open and irregular systems, such as random lasers
and lasers with chaotic resonators.

\begin{acknowledgments}

Financial support was provided by the Deutsche Forschungs\-gemein\-schaft via
the SFB/TR12 and FOR557~(O.Z.) and by PSC-CUNY via grants No.~62680-00 39 and
No.~61788-00 39~(L.D.).

\end{acknowledgments}

\appendix

\section{Derivation of Eqs.~\eq{Pq} and~\eq{nq}
\label{deriv}}

For a few lowest values of $q$ the validity of Eqs.~\eq{Pq}
and~\eq{nq} can be checked directly. To prove these relations by induction, we
assume that $\Delta n_\omega^{(q - 1)}$ is given by Eq.~\eq{nq}, then derive
$P_\omega^{(q)}$~\eq{Pq}, and, finally, obtain~$\Delta n_\omega^{(q + 1)}$.

According to Eq.~\eq{poleqft},
\eqn{
  P_\omega^{(q)} = i \hbar \gamma_\parallel  \frac A \pi D(\omega)
  \int d \omega'\, E_{\omega'} \Delta n_{\omega - \omega'}^{(q - 1)}.
}
Substituting $\Delta n_\omega^{(q - 1)}$ from Eq.~\eq{nq}, we immediately see
that the factor before the integral in Eq.~\eq{Pq} is reproduced. To calculate
the integral, we consider separately the two contributions: $\Delta n_{\omega -
\omega'}^{(q - 1)} = \wtilde {\Delta n}_{\omega - \omega'}^{(q - 1)} +
\l[\wtilde {\Delta n}_{\omega' - \omega}^{(q - 1)}\r]^*$, where $\wtilde
{\Delta n}_\omega^{(q - 1)}$ is given explicitly by Eq.~\eq{nq} and $\l[\wtilde
{\Delta n}_{- \omega}^{(q - 1)}\r]^*$ is~$\text{c.c.}(\omega \to -\omega)$.
When integrating $\wtilde {\Delta n}_{\omega - \omega'}^{(q - 1)}$ we introduce
a new variable
\eqn{
  \omega_{q - 1} = \omega' - \omega + \omega_1 - \omega_2 + \cdots - \omega_{q
  - 3} + \omega_{q - 2}.
}
Then the integral over $\omega'$ becomes
\aln{
  &\int d \omega'\, D_\parallel (\omega - \omega')\, E_{\omega'} E_{\omega' -
  \omega  + \omega_1 - \omega_2 + \cdots - \omega_{q - 3} + \omega_{q - 2}}^*
  \notag \\
  &= \int d \omega_{q - 1} D_\parallel (\omega_1 - \omega_2 + \cdots +
  \omega_{q - 2} - \omega_{q - 1}) \notag \\
  &\phantom{= \int}\times E_{\omega - \omega_1 + \omega_2 - \cdots - \omega_{q
  - 2} + \omega_{q - 1}} E_{\omega_{q - 1}}^*.
}
Comparison with Eq.~\eq{Pq} shows that $\wtilde {\Delta n}_{\omega -
\omega'}^{(q - 1)}$ contribution yields the part of the $P_\omega^{(q)}$
integrand proportional to $D(\omega_1 - \omega_2 + \cdots + \omega_{q - 2})$.
When integrating $\l[\wtilde {\Delta n}_{\omega' - \omega}^{(q - 1)}\r]^*$, we
first exchange the labels $\omega_1 \leftrightarrow \omega_2$, $\omega_3
\leftrightarrow \omega_4$, \ldots, $\omega_{q - 2} \rightarrow \omega_{q -  1}$
and then define the variable
\eqn{
  \omega_{q - 2} = \omega - \omega' - \omega_1 + \omega_2 - \cdots + \omega_{q
  - 3} + \omega_{q - 1}.
}
Transforming the integral over $\omega'$~as
\aln{
  &\int d \omega'\, D_\parallel (\omega - \omega')\, E_{\omega'} E_{\omega -
  \omega'  - \omega_1 + \omega_2 - \cdots + \omega_{q - 3} + \omega_{q - 1}}
  \notag \\
  &= \int d \omega_{q - 2} D_\parallel (\omega_1 - \omega_2 + \cdots +
  \omega_{q - 2} - \omega_{q - 1}) \notag \\
  &\phantom{= \int}\times E_{\omega - \omega_1 + \omega_2 - \cdots - \omega_{q
  - 2} + \omega_{q - 1}} E_{\omega_{q - 2}}
}
we obtain the part of the $P_\omega^{(q)}$  integrand proportional to
$D^*(\omega_2 - \omega_1 + \cdots + \omega_{q - 1})$.

Next we derive $\Delta n_\omega^{(q + 1)}$ using Eq.~\eq{inveqft},
\aln{
  \Delta n_\omega^{(q + 1)} = &\frac {-i} {\pi \hbar \gamma_\parallel}\,
  D_\parallel (\omega) \int d \omega'\, E_{\omega' - \omega}^*
  P_{\omega'}^{(q)} \notag \\
  &+ \text{c.c.}(\omega \to -\omega).
}
Clearly, the factor before the integral in Eq.~\eq{nq} follows after
substitution of $P_{\omega'}^{(q)}$~\eq{Pq}. Introducing the new variable
\eqn{
  \omega_q = \omega' - \omega_1 + \omega_2 - \cdots - \omega_{q
  - 2} + \omega_{q - 1}
}
we rewrite the $\omega'$~integral
\aln{
  &\int d \omega'\, D(\omega') \, E_{\omega' - \omega}^* E_{\omega' -  \omega_1
  + \omega_2 - \cdots - \omega_{q - 2} + \omega_{q - 1}} \notag \\
  &= \int d \omega_q\, D(\omega_1 - \omega_2 + \cdots - \omega_{q - 1} +
  \omega_q) \notag \\
  &\phantom{= \int}\times  E_{- \omega + \omega_1 - \omega_2 + \cdots -
  \omega_{q - 1} +  \omega_q}^* E_{\omega_q}.
}
This completes the proof of Eqs.~\eq{Pq} and~\eq{nq}.

\begin{widetext}
\section{General expressions for diagrams
\label{genexpr}}

The diagrams $\wtilde X_{qj}^0$ ($q = 3, 5, \ldots$) are
given by the following analytical expression:
\aln{
  \wtilde X_{qj}^0 =\, &\sum_{\wtilde X_{qj}^0} |E_{l_2} (\mbf r, t)|^2\,
  |E_{l_4} (\mbf r, t)|^2 \cdots |E_{l_{q - 1}} (\mbf r, t)|^2 \notag \\
  &\times D_\parallel (\omega_{l_1} - \omega_{l_2})\, D_\parallel (\omega_{l_1}
  - \omega_{l_2} + \omega_{l_3} - \omega_{l_4}) \cdots D_\parallel
  (\omega_{l_1} - \omega_{l_2} + \cdots + \omega_{l_{q - 2}} - \omega_{l_{q -
  1}}) \notag \\
  &\times [D(\omega_{l_1}) + D^*(\omega_{l_2})] [D(\omega_{l_1} - \omega_{l_2}
  + \omega_{l_3}) + D^*(\omega_{l_2} - \omega_{l_1} + \omega_{l_4})] \cdots
  \notag \\
  &\times [D(\omega_{l_1} - \omega_{l_2} + \omega_{l_3} - \cdots - \omega_{l_{q
  - 3}} + \omega_{l_{q - 2}}) + D^*(\omega_{l_2} - \omega_{l_1} + \omega_{l_4}
  - \cdots - \omega_{l_{q - 4}} + \omega_{l_{q - 1}})].
\label{Xqj0}
}
The symbol $\displaystyle{\sum_{\wtilde X_{qj}^0}}$ denotes a summation over
the lasing-mode indices $l_2, l_4, \ldots, l_{q - 1} = 1, \ldots, N_{\text m}$
according to the rules: (i) if the indices $l_i$ and $l_j$ are connected in the
diagram~$\wtilde X_{qj}^0$, set $l_i = l_j$ and (ii) the terms with four or
more indices $l_1, l_2, \ldots, l_q, l$ equal are excluded unless the links
connecting the affected vertices do not intersect.

We denote by $X_{qj}^0$ connected diagrams and subdiagrams containing the
vertex~$l$. With the ordering of $\wtilde X_{qj}^0$ such that the connected
diagrams come before the disconnected diagrams for a given~$q$, we can identify
$X_{qj}^0 = \wtilde X_{qj}^0$ ($j \leq N_q$), where $N_q$ is the number of
connected diagrams. The subdiagrams $X_{qj}$ without the special vertex have a
variable index~$l_{q+1}$ in place of the fixed index~$l$. The analytical
expression for these diagrams has the form
\aln{
  X_{qj} =\, &\sum_{X_{qj}} |E_{l_2} (\mbf r, t)|^2\,
  |E_{l_4} (\mbf r, t)|^2 \cdots |E_{l_{q + 1}} (\mbf r, t)|^2 \notag \\
  &\times D_\parallel (\omega_{l_1} - \omega_{l_2})\, D_\parallel (\omega_{l_1}
  - \omega_{l_2} + \omega_{l_3} - \omega_{l_4}) \cdots D_\parallel
  (\omega_{l_1} - \omega_{l_2} + \cdots + \omega_{l_{q -
  2}} - \omega_{l_{q - 1}}) \notag \\
  &\times [D(\omega_{l_1}) + D^*(\omega_{l_2})] [D(\omega_{l_1} - \omega_{l_2}
  + \omega_{l_3}) + D^*(\omega_{l_2} - \omega_{l_1} + \omega_{l_4})] \cdots
  \notag \\
  &\times [D(\omega_{l_1} - \omega_{l_2} + \omega_{l_3} - \cdots - \omega_{l_{q
  - 1}} + \omega_{l_q}) + D^*(\omega_{l_2} - \omega_{l_1} + \omega_{l_4}
  - \cdots - \omega_{l_{q - 2}} + \omega_{l_{q + 1}})].
\label{Xqj}
}
\end{widetext}
The sum $\displaystyle{\sum_{X_{qj}}}$ is defined analogously to the
sum~$\displaystyle{\sum_{\wtilde X_{qj}^0}}$ above. Note that the diagrams
$\wtilde X_{qj}^0$ and $X_{qj}^0$ are of the order~$q - 1$ in the electric
field, while the subdiagrams $X_{qj}$ are of the order~$q + 1$.

To show the multiplicativity property, let us assume that a disconnected
diagram $\wtilde X_{qj}^0$ can be cut in two, possibly disconnected
subdiagrams, $\wtilde X_{q'j'}^0$ and~$X_{q''j''}$, such that $q = q' + q'' +
1$. We will label the vertices of $X_{q''j''}$ as $l_1, \ldots, l_{q'' + 1}$
and identify them with the first $q'' + 1$ vertices of the diagram~$\wtilde
X_{qj}^0$. The last $q' + 1$ vertices of~$\wtilde X_{qj}^0$, $l_{q'' + 2},
\ldots, l_q, l$, are also the vertices of~$X_{q'j'}^0$. We need to show that,
in the sum~$\displaystyle{\sum_{\wtilde X_{qj}^0}}$, the two groups of indices
can be split between the sums $\displaystyle{\sum_{X_{q''j''}}}$
and~$\displaystyle{\sum_{X_{q'j'}^0}}$, respectively. This would mean that the
arguments of the functions $D$ and $D_\parallel$ can contain only the indices
belonging to one of the groups. This is, indeed, the case, since the arguments
having less than $q'' + 1$ frequencies contain only the indices from the
first group, while in the arguments with more frequencies the first $q'' + 1$
frequencies cancel due to the cut~as
\eqn{
  \omega_1 + \omega_3 + \cdots + \omega_{q''} = \omega_2 + \omega_4 + \cdots +
  \omega_{q'' + 1}.
}
According to this equality, the number of nontrivial factors $D_\parallel \ne
1$ in Eqs.~\eq{Xqj0} and~\eq{Xqj} is $(q - 1)/2$ minus the number of cuts (in
a disconnected diagram).



\begin{thebibliography}{24}
\expandafter\ifx\csname natexlab\endcsname\relax\def\natexlab#1{#1}\fi
\expandafter\ifx\csname bibnamefont\endcsname\relax
  \def\bibnamefont#1{#1}\fi
\expandafter\ifx\csname bibfnamefont\endcsname\relax
  \def\bibfnamefont#1{#1}\fi
\expandafter\ifx\csname citenamefont\endcsname\relax
  \def\citenamefont#1{#1}\fi
\expandafter\ifx\csname url\endcsname\relax
  \def\url#1{\texttt{#1}}\fi
\expandafter\ifx\csname urlprefix\endcsname\relax\def\urlprefix{URL }\fi
\providecommand{\bibinfo}[2]{#2}
\providecommand{\eprint}[2][]{\url{#2}}

\bibitem[{\citenamefont{Frolov et~al.}(1999)\citenamefont{Frolov, Vardeny,
  Zakhidov, and Baughman}}]{frol99}
\bibinfo{author}{\bibfnamefont{S.~V.} \bibnamefont{Frolov}},
  \bibinfo{author}{\bibfnamefont{Z.~V.} \bibnamefont{Vardeny}},
  \bibinfo{author}{\bibfnamefont{A.}~\bibnamefont{Zakhidov}}, \bibnamefont{and}
  \bibinfo{author}{\bibfnamefont{R.~H.} \bibnamefont{Baughman}},
  \bibinfo{journal}{Opt. Commun} \textbf{\bibinfo{volume}{162}},
  \bibinfo{pages}{241} (\bibinfo{year}{1999}).

\bibitem[{\citenamefont{Cao et~al.}(1999)\citenamefont{Cao, Zhao, Ho, Seelig,
  Wang, and Chang}}]{cao99}
\bibinfo{author}{\bibfnamefont{H.}~\bibnamefont{Cao}},
  \bibinfo{author}{\bibfnamefont{Y.~G.} \bibnamefont{Zhao}},
  \bibinfo{author}{\bibfnamefont{S.~T.} \bibnamefont{Ho}},
  \bibinfo{author}{\bibfnamefont{E.~W.} \bibnamefont{Seelig}},
  \bibinfo{author}{\bibfnamefont{Q.~H.} \bibnamefont{Wang}}, \bibnamefont{and}
  \bibinfo{author}{\bibfnamefont{R.~P.~H.} \bibnamefont{Chang}},
  \bibinfo{journal}{Phys. Rev. Lett.} \textbf{\bibinfo{volume}{82}},
  \bibinfo{pages}{2278} (\bibinfo{year}{1999}).

\bibitem[{\citenamefont{Fang et~al.}(2007)\citenamefont{Fang, Cao, and
  Solomon}}]{fang07}
\bibinfo{author}{\bibfnamefont{W.}~\bibnamefont{Fang}},
  \bibinfo{author}{\bibfnamefont{H.}~\bibnamefont{Cao}}, \bibnamefont{and}
  \bibinfo{author}{\bibfnamefont{G.~S.} \bibnamefont{Solomon}},
  \bibinfo{journal}{Appl. Phys. Lett.} \textbf{\bibinfo{volume}{90}},
  \bibinfo{pages}{081108} (\bibinfo{year}{2007}).

\bibitem[{\citenamefont{Sargent~III et~al.}(1974)\citenamefont{Sargent~III,
  Scully, and Lamb}}]{sarg74}
\bibinfo{author}{\bibfnamefont{M.}~\bibnamefont{Sargent~III}},
  \bibinfo{author}{\bibfnamefont{M.~O.} \bibnamefont{Scully}},
  \bibnamefont{and} \bibinfo{author}{\bibfnamefont{W.~E.} \bibnamefont{Lamb},
  \bibfnamefont{Jr.}}, \emph{\bibinfo{title}{Laser Physics}}
  (\bibinfo{publisher}{Addison-Wesley}, \bibinfo{address}{Reading},
  \bibinfo{year}{1974}).

\bibitem[{\citenamefont{Haken}(1984)}]{hake84}
\bibinfo{author}{\bibfnamefont{H.}~\bibnamefont{Haken}},
  \emph{\bibinfo{title}{Laser Theory}} (\bibinfo{publisher}{Springer},
  \bibinfo{address}{Berlin}, \bibinfo{year}{1984}).

\bibitem[{\citenamefont{{Fox} and {Li}}(1968)}]{fox68}
\bibinfo{author}{\bibfnamefont{A.}~\bibnamefont{{Fox}}} \bibnamefont{and}
  \bibinfo{author}{\bibfnamefont{T.}~\bibnamefont{{Li}}},
  \bibinfo{journal}{IEEE Journal of Quantum Electronics}
  \textbf{\bibinfo{volume}{4}}, \bibinfo{pages}{460} (\bibinfo{year}{1968}).

\bibitem[{\citenamefont{Siegman}(1989{\natexlab{a}})}]{sieg89a}
\bibinfo{author}{\bibfnamefont{A.~E.} \bibnamefont{Siegman}},
  \bibinfo{journal}{Phys. Rev. A} \textbf{\bibinfo{volume}{39}},
  \bibinfo{pages}{1253} (\bibinfo{year}{1989}{\natexlab{a}}).

\bibitem[{\citenamefont{Siegman}(1989{\natexlab{b}})}]{sieg89b}
\bibinfo{author}{\bibfnamefont{A.~E.} \bibnamefont{Siegman}},
  \bibinfo{journal}{Phys. Rev. A} \textbf{\bibinfo{volume}{39}},
  \bibinfo{pages}{1264} (\bibinfo{year}{1989}{\natexlab{b}}).

\bibitem[{\citenamefont{{Moiseyev}}(1998)}]{mois98}
\bibinfo{author}{\bibfnamefont{N.}~\bibnamefont{{Moiseyev}}},
  \bibinfo{journal}{Physics Reports} \textbf{\bibinfo{volume}{302}},
  \bibinfo{pages}{212} (\bibinfo{year}{1998}).

\bibitem[{\citenamefont{Ching et~al.}(1998)\citenamefont{Ching, Leung, Maassen
  van~den Brink, Suen, Tong, and Young}}]{chin98}
\bibinfo{author}{\bibfnamefont{E.~S.~C.} \bibnamefont{Ching}},
  \bibinfo{author}{\bibfnamefont{P.~T.} \bibnamefont{Leung}},
  \bibinfo{author}{\bibfnamefont{A.}~\bibnamefont{Maassen van~den Brink}},
  \bibinfo{author}{\bibfnamefont{W.~M.} \bibnamefont{Suen}},
  \bibinfo{author}{\bibfnamefont{S.~S.} \bibnamefont{Tong}}, \bibnamefont{and}
  \bibinfo{author}{\bibfnamefont{K.}~\bibnamefont{Young}},
  \bibinfo{journal}{Rev. Mod. Phys.} \textbf{\bibinfo{volume}{70}},
  \bibinfo{pages}{1545} (\bibinfo{year}{1998}).

\bibitem[{\citenamefont{Dutra and Nienhuis}(2000)}]{dutr00}
\bibinfo{author}{\bibfnamefont{S.~M.} \bibnamefont{Dutra}} \bibnamefont{and}
  \bibinfo{author}{\bibfnamefont{G.}~\bibnamefont{Nienhuis}},
  \bibinfo{journal}{Phys. Rev. A} \textbf{\bibinfo{volume}{62}},
  \bibinfo{pages}{063805} (\bibinfo{year}{2000}).

\bibitem[{\citenamefont{Viviescas and Hackenbroich}(2003)}]{vivi03}
\bibinfo{author}{\bibfnamefont{C.}~\bibnamefont{Viviescas}} \bibnamefont{and}
  \bibinfo{author}{\bibfnamefont{G.}~\bibnamefont{Hackenbroich}},
  \bibinfo{journal}{Phys. Rev. A} \textbf{\bibinfo{volume}{67}},
  \bibinfo{pages}{013805} (\bibinfo{year}{2003}).

\bibitem[{\citenamefont{Viviescas and Hackenbroich}(2004)}]{vivi04}
\bibinfo{author}{\bibfnamefont{C.}~\bibnamefont{Viviescas}} \bibnamefont{and}
  \bibinfo{author}{\bibfnamefont{G.}~\bibnamefont{Hackenbroich}},
  \bibinfo{journal}{J. Opt. B: Quantum Semiclass. Opt.}
  \textbf{\bibinfo{volume}{6}}, \bibinfo{pages}{211} (\bibinfo{year}{2004}).

\bibitem[{\citenamefont{Morse and Feshbach}(1953)}]{mors53}
\bibinfo{author}{\bibfnamefont{P.~M.} \bibnamefont{Morse}} \bibnamefont{and}
  \bibinfo{author}{\bibfnamefont{H.}~\bibnamefont{Feshbach}},
  \emph{\bibinfo{title}{Methods of Theoretical Physics}},
  vol.~\bibinfo{volume}{1} (\bibinfo{publisher}{McGrow-Hill},
  \bibinfo{address}{New York}, \bibinfo{year}{1953}).

\bibitem[{\citenamefont{T{\"u}reci et~al.}(2006)\citenamefont{T{\"u}reci,
  Stone, and Collier}}]{ture06}
\bibinfo{author}{\bibfnamefont{H.~E.} \bibnamefont{T{\"u}reci}},
  \bibinfo{author}{\bibfnamefont{A.~D.} \bibnamefont{Stone}}, \bibnamefont{and}
  \bibinfo{author}{\bibfnamefont{B.}~\bibnamefont{Collier}},
  \bibinfo{journal}{Phys. Rev. A} \textbf{\bibinfo{volume}{74}},
  \bibinfo{pages}{043822} (\bibinfo{year}{2006}).

\bibitem[{\citenamefont{Deych}(2005{\natexlab{a}})}]{deyc05a}
\bibinfo{author}{\bibfnamefont{L.~I.} \bibnamefont{Deych}},
  \bibinfo{journal}{Phys. Rev. Lett.} \textbf{\bibinfo{volume}{95}},
  \bibinfo{pages}{043902} (\bibinfo{year}{2005}{\natexlab{a}}).

\bibitem[{\citenamefont{Deych}(2005{\natexlab{b}})}]{deyc05b}
\bibinfo{author}{\bibfnamefont{L.}~\bibnamefont{Deych}}, in
  \emph{\bibinfo{booktitle}{Complex Mediums VI: Light and Complexity}}, edited
  by \bibinfo{editor}{\bibfnamefont{M.~W.} \bibnamefont{McCall}},
  \bibinfo{editor}{\bibfnamefont{G.}~\bibnamefont{Dewar}}, \bibnamefont{and}
  \bibinfo{editor}{\bibfnamefont{M.~A.} \bibnamefont{Noginov}}
  (\bibinfo{publisher}{SPIE}, \bibinfo{year}{2005}{\natexlab{b}}), vol.
  \bibinfo{volume}{5924}, p. \bibinfo{pages}{59240B}.

\bibitem[{\citenamefont{Misirpashaev and Beenakker}(1998)}]{misi98}
\bibinfo{author}{\bibfnamefont{T.~S.} \bibnamefont{Misirpashaev}}
  \bibnamefont{and} \bibinfo{author}{\bibfnamefont{C.~W.~J.}
  \bibnamefont{Beenakker}}, \bibinfo{journal}{Phys. Rev. A}
  \textbf{\bibinfo{volume}{57}}, \bibinfo{pages}{2041} (\bibinfo{year}{1998}).

\bibitem[{\citenamefont{Hackenbroich}(2005)}]{hack05}
\bibinfo{author}{\bibfnamefont{G.}~\bibnamefont{Hackenbroich}},
  \bibinfo{journal}{J. Phys. A} \textbf{\bibinfo{volume}{38}},
  \bibinfo{pages}{10537} (\bibinfo{year}{2005}).

\bibitem[{\citenamefont{Zaitsev et~al.}(2009)\citenamefont{Zaitsev, Deych, and
  Shuvayev}}]{zait09}
\bibinfo{author}{\bibfnamefont{O.}~\bibnamefont{Zaitsev}},
  \bibinfo{author}{\bibfnamefont{L.}~\bibnamefont{Deych}}, \bibnamefont{and}
  \bibinfo{author}{\bibfnamefont{V.}~\bibnamefont{Shuvayev}},
  \bibinfo{journal}{Phys. Rev. Lett.} \textbf{\bibinfo{volume}{102}},
  \bibinfo{pages}{043906} (\bibinfo{year}{2009}).

\bibitem[{\citenamefont{T{\"u}reci et~al.}(2008)\citenamefont{T{\"u}reci, Ge,
  Rotter, and Stone}}]{ture08}
\bibinfo{author}{\bibfnamefont{H.~E.} \bibnamefont{T{\"u}reci}},
  \bibinfo{author}{\bibfnamefont{L.}~\bibnamefont{Ge}},
  \bibinfo{author}{\bibfnamefont{S.}~\bibnamefont{Rotter}}, \bibnamefont{and}
  \bibinfo{author}{\bibfnamefont{A.~D.} \bibnamefont{Stone}},
  \bibinfo{journal}{Science} \textbf{\bibinfo{volume}{320}},
  \bibinfo{pages}{643} (\bibinfo{year}{2008}), \bibinfo{note}{see also
  Supporting Online Material.}

\bibitem[{\citenamefont{{T{\"u}reci} et~al.}(2009)\citenamefont{{T{\"u}reci},
  {Stone}, {Ge}, {Rotter}, and {Tandy}}}]{ture09}
\bibinfo{author}{\bibfnamefont{H.~E.} \bibnamefont{{T{\"u}reci}}},
  \bibinfo{author}{\bibfnamefont{A.~D.} \bibnamefont{{Stone}}},
  \bibinfo{author}{\bibfnamefont{L.}~\bibnamefont{{Ge}}},
  \bibinfo{author}{\bibfnamefont{S.}~\bibnamefont{{Rotter}}}, \bibnamefont{and}
  \bibinfo{author}{\bibfnamefont{R.~J.} \bibnamefont{{Tandy}}},
  \bibinfo{journal}{Nonlinearity} \textbf{\bibinfo{volume}{22}},
  \bibinfo{pages}{1} (\bibinfo{year}{2009}).

\bibitem[{\citenamefont{{T{\"u}reci} et~al.}(2007)\citenamefont{{T{\"u}reci},
  {Stone}, and {Ge}}}]{ture07}
\bibinfo{author}{\bibfnamefont{H.~E.} \bibnamefont{{T{\"u}reci}}},
  \bibinfo{author}{\bibfnamefont{A.~D.} \bibnamefont{{Stone}}},
  \bibnamefont{and} \bibinfo{author}{\bibfnamefont{L.}~\bibnamefont{{Ge}}},
  \bibinfo{journal}{Phys. Rev. A} \textbf{\bibinfo{volume}{76}},
  \bibinfo{pages}{013813} (\bibinfo{year}{2007}).

\bibitem[{\citenamefont{Shuvayev and Deych}(2009)}]{shuv09}
\bibinfo{author}{\bibfnamefont{V.}~\bibnamefont{Shuvayev}} \bibnamefont{and}
  \bibinfo{author}{\bibfnamefont{L.}~\bibnamefont{Deych}},
  \bibinfo{journal}{unpublished}  (\bibinfo{year}{2009}).

\end{thebibliography}

\end{document}